\documentclass{article}

\usepackage{arxiv}
\usepackage[utf8]{inputenc}
\usepackage[inline]{enumitem}

\usepackage[dvipsnames]{xcolor}
\usepackage{tabularx}
\usepackage{tabulary}
\usepackage{makecell}

\usepackage{multirow}
\usepackage{subcaption}
\usepackage{booktabs}
\usepackage{graphicx}
\usepackage{pifont}
\usepackage{threeparttable}
\usepackage[format=plain, labelfont={bf}, textfont=it]{caption}
\usepackage[hidelinks]{hyperref}
\usepackage{amsmath}
\usepackage[acronym, shortcuts, nohypertypes={acronym}]{glossaries}
\usepackage{todonotes}
\usepackage{tcolorbox}
\usepackage{hyperref}

\usepackage[natbib, bibencoding=utf8, citestyle=numeric, bibstyle=ieee, maxbibnames=999, maxcitenames=2, mincitenames=1, sortcites]{biblatex}
\bibliography{bibliography.bib}

\usepackage[capitalise, nameinlink, noabbrev]{cleveref}

\newcommand\myshade{75}
\colorlet{myurlcolor}{NavyBlue}
\hypersetup{
  colorlinks=true,
  linkcolor=black,
  citecolor=black,
  urlcolor=myurlcolor!\myshade!black,
}

\newacronym{AI}{AI}{artificial intelligence}
\newacronym{ASR}{ASR}{automatic speech recognition}
\newacronym{CCC}{CCC}{concordance correlation coefficient}
\newacronym{CNN}{CNN}{convolutional neural network}
\newacronym{NLP}{NLP}{natural language processing}
\newacronym{SER}{SER}{speech emotion recognition}
\newacronym{SGD}{SGD}{stochastic gradient descent}
\newacronym{SNR}{SNR}{signal-to-noise ratio}
\newacronym{UAR}{UAR}{unweighted average recall}
\newacronym{WAR}{WAR}{weighted average recall}

\newcommand\cnn{\mbox{\emph{CNN14}}}
\newcommand\wbase{\mbox{\emph{w2v2-b}}}
\newcommand\hbase{\mbox{\emph{hubert-b}}}
\newcommand\wlarge{\mbox{\emph{w2v2-L}}}
\newcommand\hlarge{\mbox{\emph{hubert-L}}}
\newcommand\wrobust{\mbox{\emph{w2v2-L-robust}}}
\newcommand\wrobustpruned{\mbox{\emph{w2v2-L-robust-12}}}
\newcommand\wvox{\mbox{\emph{w2v2-L-vox}}}
\newcommand\wxlsr{\mbox{\emph{w2v2-L-xls-r}}}
\newcommand\wwopre{\mbox{\emph{w2v2-L-w/o-pretrain}}}

\newcommand{\review}[1]{\textcolor{black}{#1}}

\newcommand\release{\mbox{\url{https://github.com/audeering/w2v2-how-to}}}

\newcommand\msppodcast{\mbox{MSP-Podcast}}
\newcommand\iemocap{\mbox{IEMOCAP}}
\newcommand\mosi{\mbox{MOSI}}

\newcommand{\wtov}{wav2vec\,2.0}
\newcommand{\hubert}{HuBERT}

\newcommand{\eg}{e.\,g.\ }
\newcommand{\ie}{i.\,e.\ }
\newcommand{\wrt}{w.\,r.\,t.\ }
\newcommand{\cf}{{cf.\ }}

\renewcommand{\check}{\ding{51}}

\title{
Dawn of the transformer era in speech emotion recognition: closing the valence gap
\thanks{\textit{\underline{Citation}}: 
\textbf{Wagner, J., Triantafyllopoulos, A., Wierstorf, H., Schmitt, M., Burkhardt, F., Eyben, F., \& Schuller, B. W. (2023). Dawn of the transformer era in speech emotion recognition: closing the valence gap. IEEE Transactions on Pattern Analysis and Machine Intelligence. 10.1109/TPAMI.2023.3263585}} 
}
\renewcommand{\shorttitle}{
Dawn of the transformer era in speech emotion recognition
}
\author{
Johannes Wagner\textsuperscript{1}, Andreas Triantafyllopoulos\textsuperscript{2}, Hagen Wierstorf\textsuperscript{1}, \textbf{Maximilian Schmitt\textsuperscript{1}}, \\\textbf{Felix Burkhardt\textsuperscript{1}}, \textbf{Florian Eyben\textsuperscript{1}}, \textbf{Bj\"{o}rn W. Schuller\textsuperscript{1,2,3}}\\
\textsuperscript{1} audEERING GmbH, Gilching, Germany\\
\textsuperscript{2} EIHW, University of Augsburg, Augsburg, Germany\\
\textsuperscript{3} GLAM, Imperial College, London, UK\\
}

\renewcommand{\headeright}{Preprint}
\renewcommand{\undertitle}{}

\newlength{\boxarc}
\newlength{\boxboxrule}
\begin{document}
\maketitle

\begin{abstract}
    Recent advances in transformer-based architectures have shown promise in several machine learning tasks.
    In the audio domain, such architectures have been successfully utilised in the field of speech emotion recognition (SER).
    However, existing works have not evaluated the influence of \emph{model size} and \emph{pre-training data} on downstream performance, and have shown limited attention to \emph{generalisation}, \emph{robustness}, \emph{fairness}, and \emph{efficiency}.
    The present contribution conducts a thorough analysis of these aspects on several pre-trained variants of {\wtov} and {\hubert} that we fine-tuned on the dimensions arousal, dominance, and valence of {\msppodcast}, while additionally using {\iemocap} and {\mosi} to test cross-corpus generalisation.
    To the best of our knowledge, we obtain the top performance for valence prediction without use of explicit linguistic information, with a concordance correlation coefficient (CCC) of $.638$ on {\msppodcast}.
    Our investigations reveal that transformer-based architectures are more robust  compared to a CNN-based baseline and fair with respect to gender groups, but not towards individual speakers.
    Finally, we show that their success on valence is based on implicit linguistic information, which explains why they perform on-par with recent multimodal approaches that explicitly utilise textual information.
    To make our findings reproducible, we release the best performing model to the community.
\end{abstract}

\glsresetall

\keywords{Affective Computing \and Speech Emotion Recognition \and Transformers.}

\section{Introduction}
\label{sec:introduction}

Automatic \ac{SER} is a key enabling technology for 
facilitating better human-to-machine interactions~\citep{Schuller18-SER}.
\ac{SER} research is dominated by two conceptual paradigms: \emph{discrete emotions}~\citep{ekman1992argument} and emotional \emph{dimensions}~\citep{russell1977evidence}.
The first investigates emotional categories like \emph{happy} or \emph{sad}, while the latter focuses on the dimensions of \emph{arousal}, \emph{valence}, and \emph{dominance}~\citep{russell1977evidence}.

A \ac{SER} system achieves this through the linguistic (\emph{what} has been said) or the paralinguistic (\emph{how} it has been said) stream~\citep{Schuller18-SER, atmaja2022survey, pantic2009survey}.
The linguistic stream is better suited for valence recognition~\citep{calvo2010affect, sewa2021}
and can draw from recent advances in \ac{ASR} and \ac{NLP}~\citep{sahu2019multi},
but might be limited to a single language. 
Paralinguistics works better for arousal and dominance~\citep{calvo2010affect, sewa2021}
and has the potential to generalise across different languages.
Both paradigms
can be combined in \emph{bimodal} architectures~\citep{atmaja2022survey},
which require to execute several different models.
Instead, 
we aim towards a model that only implicitly utilises the linguistic information stream during deployment,
and does not require access to \ac{ASR} and \ac{NLP} frontends.

Although the field has seen tremendous progress in the last decades~\citep{Schuller18-SER},
three major challenges remain for real-world paralinguistics-based \ac{SER} applications:
a)
improving on its inferior valence performance~\citep{sewa2021, triantafyllopoulos2021multistage},
b)
overcoming issues of generalisation and robustness~\citep{Oates19-RSE, Triantafyllopoulos19-TRS}, and
c)
alleviating individual- and group-level fairness concerns, which is a prerequisite for ethical emotion recognition technology~\citep{batliner2020ethics, cheong2021hitchhiker}.
Previous works have attempted to tackle these issues in isolation,
but combining them is not straightforward.

In recent years,
the \ac{AI} field is undergoing a major paradigm shift,
moving from specialised architectures trained for a given task
to general-purpose \emph{foundation models}
that can be adapted to several use-cases~\citep{bommasani2021opportunities}.
Such models have seen tremendous success in computer vision~\citep{chen2020simple, han2022survey},
\ac{NLP}~\citep{vaswani2017attention},
and computer audition~\citep{baevski2020wav2vec, hsu2021hubert},
including \ac{SER}~\citep{wang2021finetuned, Latif21-SOD}.
Among others,
{\wtov}~\citep{baevski2020wav2vec}
and {\hubert}~\citep{hsu2021hubert}
have emerged as foundation model candidates for speech-related applications.
We evaluate several publicly-available pre-trained
variants of those models
for dimensional \ac{SER},
and show that they can achieve state-of-the art results for valence.
We further analyze the influence of the model architecture, the 
\review{pre-}training data,
how well the models generalise,
their robustness,
fairness,
and efficiency.
Moreover,
we make our best performing model publicly available~\citep{wagner2022model}.
To our best knowledge
this is the first transformer-based dimensional \ac{SER} model released to the community.
For an introduction on how to use it,
please visit: {\release}.

The remainder of this paper is organised as follows.
\cref{sec:related} discusses related work,
\cref{sec:setup} presents the models, databases, and evaluation methods.
\cref{sec:evaluation} shows the results 
\review{and investigates why transformer models are able to close the valence gap and improve performance with respect to robustness and fairness.}
\cref{sec:efficiency} investigates efficiency improvements,
before \cref{sec:summary} summarises the results,
and \cref{sec:conclusion} concludes the paper.


\section{Related Work}
\label{sec:related}

\begin{table}[!ht]
    \centering
    \caption{
        State-of-the-art 4-class emotion recognition performance on {\iemocap}
        using transformer-based architectures ranked by \acf{UAR} / \acf{WAR}.
        The table encodes whether the base (b) or large (L) architecture was used
        as well as whether the pre-trained model was fine-tuned for speech recognition (FT-SR).
        The column FT-D marks if the transformer layers were further fine-tuned
        during the down-stream classification task.
    }
    \begin{threeparttable}
        \begin{tabular}{llcccccc}
            \toprule
              & \thead[l]{Work} & \thead{Model} & \thead{L} & \thead{FT-SR} & \thead{FT-D} & \thead{\acs{UAR}} & \thead{\acs{WAR}}  \\
            \midrule
            1 & \citep{krishna2021using} & \emph{\wlarge} & \check & ~ & ~ & 60.0 & \\
            2 & \citep{yuan2021role}\tnote{*} & \emph{\wlarge} & \check & ~ & ~ & 62.5 & 62.6 \\
            3 & \citep{wang2021finetuned} & \emph{\wbase} & ~ & ~ & ~ & & 63.4 \\
            4 & \citep{yang2021superb} & \emph{\wbase} & ~ & ~ & ~ & 63.4 & \\
            5 & \citep{pepino2021emotion} & \emph{\wbase} & ~ & \check & ~ & 63.8 & \\
            6 & \citep{wang2021finetuned} & \emph{\hbase} & ~ & ~ & ~ & & 64.9 \\
            7 & \citep{yang2021superb} & \emph{\hbase} & ~ & ~ & ~ & 64.9 & \\
            8 & \citep{wang2021finetuned} & \emph{\wlarge} & \check & ~ & ~ & & 65.6 \\
            9 & \citep{yang2021superb} & \emph{\wlarge} & \check & ~ & ~ & 65.6 & \\
            10 & \citep{pepino2021emotion} & \emph{\wbase} & ~ & ~ & ~ & 67.2 & \\
            11 & \citep{wang2021finetuned} & \emph{\hlarge} & \check & ~ & ~ & & 67.6 \\
            12 & \citep{yang2021superb} & \emph{\hlarge} & \check & ~ & ~ & 67.6 & \\
            13 & \citep{chen2021exploring} & \emph{\wbase} & ~ & ~ & \check & 69.9 & \\
            14 & \citep{makiuchi2021multimodal} & \emph{\wlarge} & \check & ~ & ~ & 70.7 & \\
            15 & \citep{wang2021finetuned} & \emph{\wbase} & ~ & \check & \check & & 73.8 \textcolor{ForestGreen}{(68.3)}\tnote{**}  \\
            16 & \citep{chen2021exploring} & \emph{\wbase} & ~ & ~ & \check & 74.3  & \\
            17 & \citep{wang2021finetuned} & \emph{\hbase} & ~ & ~ & \check & & 76.6 \textcolor{ForestGreen}{(69.7)}\tnote{**}  \\
            18 & \citep{wang2021finetuned} & \emph{\wlarge} & \check & \check & \check & & 76.8 \textcolor{ForestGreen}{(69.1)}\tnote{**}  \\
            19 & \citep{wang2021finetuned} & \emph{\wbase} & ~ & ~ & \check & & 77.0 \textcolor{ForestGreen}{(71.0)}\tnote{**}  \\
            20 & \citep{wang2021finetuned} & \emph{\wlarge} & \check & ~ & \check & & 77.5 \textcolor{ForestGreen}{(71.0)}\tnote{**} \\
            21 & \citep{wang2021finetuned} & \emph{\hlarge} & \check & \check & \check & & 79.0 \textcolor{ForestGreen}{(73.0)}\tnote{**}  \\
            22 & \citep{wang2021finetuned} & \emph{\hlarge} & \check & ~ & \check & & 79.6 \textcolor{ForestGreen}{(73.0)}\tnote{**} \\
            \bottomrule
        \end{tabular}
        \begin{tablenotes}
            \item[*] For a fair comparison
            we report the result on the utterance level.
            Authors report better performance on the phonetic level.
            \item[**] \textcolor{ForestGreen}{We updated the table with their speaker independent results in parenthesis. Thanks to Dr. Leyuan Qu for the correction.}
      \end{tablenotes}
    \end{threeparttable}
    \label{tab:related-work-egemaps}
\end{table}

\review{The focus of our work is the recognition of emotional dimensions.
However, 
most related studies target emotional categories.
Since the approaches are closely related,
we consider both in this section.}

In \cref{tab:related-work-egemaps},
we provide a summary of recent works based on {\wtov} and {\hubert}
on the {\iemocap} dataset~\citep{busso2008iemocap},
on which most prior works have focused.
Results are ranked by \acf{UAR} / \acf{WAR}
on the four emotional categories of
anger (1103 utterances),
happiness (+ excitement) (1636),
sadness (1084),
and neutral (1708),
which is the typical categorical \ac{SER} formulation for {\iemocap}.
Most of the works apply leave-one-session-out cross validation (5 folds),
except \citet{yuan2021role},
using leave-one-speaker-out cross validation (10 folds),
and \citet{wang2021finetuned},
who do not explicitly mention which folds they used.
Even though
authors have used different head architectures and training procedures in their studies,
we can draw some general observations:

\begin{enumerate}
    \item Fine-tuning pre-trained weights yields a $10$\% boost.
    \item Additional \ac{ASR} fine-tuning does not help with \ac{SER} (\eg row $15$ vs row $19$ $-3.2$\%).
    \item The large architecture is typically better than the base one (\eg row $17$ vs row $22$ $+3.0$\%),
    but differences can be quite small (\eg row $19$ vs row $20$ $+.5$\% ).
    \item {\hubert} outperforms {\wtov} (\eg row $22$ vs row $20$: $+2.1$\%).
    \item When performing a fine-tuning of the transformer layers,
    a simple average pooling in combination with a linear classifier built over {\wtov} or {\hubert}
    as proposed by \citet{wang2021finetuned}
    seems sufficient and shows best performance in the ranking.
    However, some of the more complex models
    like the cross-representation encoder-decoder model proposed by \citet{makiuchi2021multimodal}
    only report results without fine-tuning the pre-trained model during the down-stream task.
\end{enumerate}

While the aforementioned studies have focused on emotional categories,
there also exist several ones
which concentrate on dimensions.
The most comparable to ours is that of \citet{srinivasan2021representation},
who fine-tuned {\wtov} / {\hubert} on arousal, dominance, and valence.
Their results show that pre-trained models are particularly good in predicting valence.
When additionally joining audio embeddings from the fine-tuned models
and text representations obtained with a pre-trained BERT model,
they got a \acf{CCC} for valence of $.683$
on the {\msppodcast} corpus~\citep{lotfian2019msppodcast}.
Furthermore, they were able to distill the multi-model system
to an audio-only model using student-teacher transfer learning,
while still reaching a \ac{CCC} of $.627$
(a massive improvement compared to the previous state-of-the-art performance of only $.377$~\citep{li2021contrastive}).
\review{
However, 
this improvement was the result of cross-modal transfer learning, 
and it remains unclear whether speech-based architectures
are by themselves able
to reach such performance level
-- a fact we further explore in our work.
}

The presented results demonstrate the great potential of {\wtov} and {\hubert} for emotion recognition. 
However, the influence of pre-training data quantity and domain remains unclear.
For instance, even though the large model shows consistently better performance,
it is unclear if that can be attributed to the additional layers
or to an $60$ fold increase of training data compared to the base model. 
Likewise there is little understanding on the impact of language,
as previous work focused in pre-training on English speech data.
In this contribution,
we present a systematic comparison of different models
pre-trained under various conditions
(\eg including noisy speech)
and evaluate them on several datasets
(in-domain and cross-corpus).

Moreover, it is important to show
that \ac{SER} models work well under noisy conditions.
\citet{jaiswal2021robustness,pappagari2020robustness,Oates19-RSE,Triantafyllopoulos19-TRS} have shown
that previous \ac{SER} models suffer from robustness issues.
We systematically investigate robustness of transformer-based models
against a variety of augmentations
\review{
that 
do not change the human perception
of the underlying emotion~\citep{jaiswal2021robustness}.
}

Finally,
we consider fairness an important,
but challenging topic for machine learning models.
Discussions in the speech processing community focus mainly on group fairness,
\eg gender~\citep{rajan2021fairness}. 
For \ac{SER} models,
only a few evaluations are available.
\citet{gorrostieta2019gender} found a decrease in \ac{CCC} for females
compared to males for arousal in {\msppodcast} (v1.3) of $.234$.
Besides group fairness,
this contribution investigates individual fairness
by estimating the influence of the speaker on the model performance,
which is a known problem for speaker verification models~\citep{doddington1998sheep}.


\section{Experimental setup}
\label{sec:setup}


\subsection{Pre-trained models}
\label{subsec:models}

\begin{table}[t]
    \centering
    \caption{
        Transformer-based models included in this study
        and details on the data used during pre-training.
        Models comprised of two architecture designs ({\wtov} and {\hubert}),
        each with two different variants (base and large).
        For each model, we list included dataset(s),
        total number of hours (h),
        number of languages (\emph{eng} if only English),
        and covered domains
        (\textbf{R}ead speech, \textbf{T}elephone conversions, \textbf{P}arliamentary speech, \textbf{Y}outube).
    }
    \scriptsize
    \begin{threeparttable}
    \begin{tabular}{lllccc}
        \toprule
        \textbf{Model} & \textbf{Datasets} & \textbf{h} & \textbf{Lang} & \textbf{Domain} \\
        \midrule
        \emph{\wbase}~\citep{baevski2020wav2vec} & LibriSpeech & 960 & eng & R \\
        \emph{\hbase}~\citep{hsu2021hubert} & LibriSpeech & 960 & eng & R \\
        \emph{\wlarge}~\citep{baevski2020wav2vec} & \textcolor{ForestGreen}{LibriSpeech}\tnote{*} & 60k & eng & R \\
        \emph{\hlarge}~\citep{hsu2021hubert} & Libri-Light & 60k & eng & R \\
        \emph{\wrobust}~\citep{hsu2021robust}  & \makecell[lt]{Libri-Light (60k)\\Fisher (2k)\\CommonVoice (700)\\Switchboard (300)} & 63k     & eng & R, T \\
        \emph{\wvox}~\citep{wang2021voxpopuli} & VoxPopuli & 100k & 23 & P \\
        \emph{\wxlsr}~\citep{babu2021xls-r} & \makecell[lt]{VoxPopuli (372k)\\ ML LibriSpeech (50k)\\CommonVoice (7k)\\ VoxLingua107 (6.6k)\\BABEL (1k)} & 436k & 128 & R, T, P, Y \\
        \bottomrule
    \end{tabular}
    \begin{tablenotes}
            \item[*] \textcolor{ForestGreen}{In our initial version, we erroneously specified Libri-Light as the training set of {\wlarge}. However, it was actually trained on LibriSpeech, same as {\wbase}.}
      \end{tablenotes}
    \end{threeparttable}
    \label{tab:models} 
\end{table}

Throughout the paper,
we discuss results obtained with transformer-based models
pre-trained on large amounts of unlabelled data. 
We investigate two main variants: {\wtov} \cite{baevski2020wav2vec} and {\hubert} \cite{hsu2021hubert}.
\review{The network architecture of both models is the same. As input, it expects a raw waveform 
normalised to have zero mean and unit variance, which is fed into a \emph{feature encoder} consisting of $7$ convolutional layers that extracts \emph{feature vectors} over time, with a dimensionality of $512$ and a step size of $20$\,ms. These features are projected to a higher dimension ($768$ or $1024$ hidden units, see below) and then fed into the \emph{encoder}. The \emph{encoder} is a series of \emph{transformer layers}, each of them consisting of a \emph{multi-head self-attention module} and several \emph{fully-connected layers}. In order to inject temporal information, the output of a \emph{convolutional layer} is added at the input of the \emph{encoder}.}

\review{The only difference between the main variants is the way they are pre-trained on unlabelled data.
In {\wtov}, the features of a certain ratio of time steps are masked, by replacing them with a learnt fixed feature vector at the input of the \emph{encoder}. A \emph{contrastive loss} between the \emph{encoder outputs} and a \emph{quantised version} of the input features is then minimised~\cite{baevski2020wav2vec}. In order to avoid learning too simple representations, the quantisation is done using a codebook, whose \emph{diversity loss} is minimised as well.}
\review{In contrast, {\hubert} minimises a \emph{cross-entropy loss} for the masked time steps, where the targets are not trained simultaneously with the model.
The pre-training is performed in several steps, where in the first step, clusters obtained by \emph{k-means clustering of MFCCs} are employed as targets and in later steps, clusters of the outputs of certain transformer layers are taken into account~\cite{hsu2021hubert}.
In following these strategies, the models try to learn the \emph{structure} of speech, resulting in a reduced need for labelled task-specific training data.}

Both {\wtov} and {\hubert} exist in two forms:
a base architecture with $12$ transformer layers of $768$ hidden units each ($95$M parameters),
and a large architecture with $24$ transformer layers of $1024$ hidden units each ($317$M parameters). 
Apart from that, we further distinguish them by the data used for pre-training.
We included the four models found in previous work (\cf \cref{sec:related}),
which are pre-trained on English audiobooks,
namely
wav2vec2-base (\wbase),
hubert-base-ls960 (\hbase),
wav2vec2-large (\wlarge),
hubert-large-ll60k (\hlarge);
the wav2vec2-large-robust model (\wrobust),
additionally trained on telephone speech;
the wav2vec2-large-100k-voxpopuli model (\wvox),
trained only on parliamentary speech in multiple languages;
and the wav2vec2-xls-r-300m model (\wxlsr),
trained on more than $400$k hours across all domains and multiple languages.
Compare \cref{tab:models} for citations and an overview of the included data.
We did not include models fine-tuned on speech recognition
as previous work showed that this does not lead to better performance.
Also note that we refer to their fine-tuned versions when we report results (\cf \cref{subsec:architecture}).


\subsection{Architecture}
\label{subsec:architecture}

\begin{figure}[t]
    \centering
    \includegraphics[width=\columnwidth]{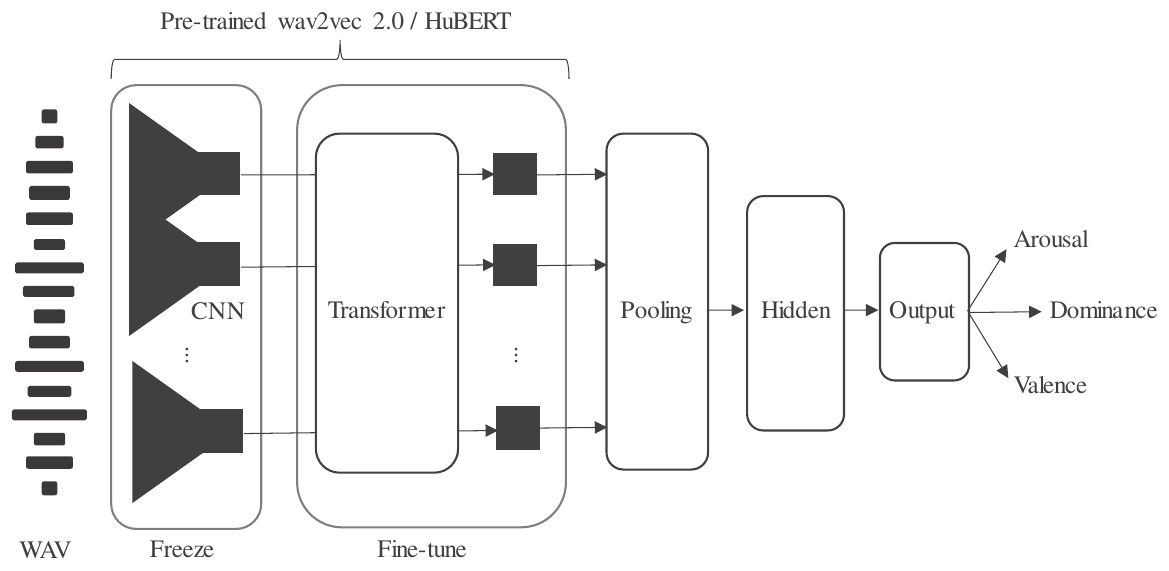}
    \caption{Proposed architecture built on {\wtov} / {\hubert}.}
    \label{fig:architecture}
\end{figure}

Inspired by \citet{wang2021finetuned}
we apply average pooling over the hidden states of the last transformer layer
and feed the result through a hidden layer and a final output layer. 
For fine-tuning on the downstream task,
we use the ADAM optimiser with \ac{CCC} loss,
which is the standard loss function
used for dimensional \ac{SER}~\citep{Trigeorgis16-AFE, li2021contrastive, triantafyllopoulos2021multistage},
and a fixed learning rate of $1\mathrm{e}{-4}$.
We run for $5$ epochs with a batch size of $32$
and keep the checkpoint with best performance on the development set. 

During training,
we freeze the CNN layers but fine-tune the transformer ones. 
According to \citet{wang2021finetuned},
such a partial fine-tuning yields better results.
When using the term fine-tuning,
we will henceforth refer to this partial fine-tuning.
These models are trained using a single random seed,
for which the performance is reported. 

We compare results to a $14$-layer Convolutional Neural Network ({\cnn}) as a standard baseline
we have been using for \ac{SER} in previous work~\citep{triantafyllopoulos2021role, triantafyllopoulos2021multistage}. 
It follows the architecture proposed by \citet{kong2019panns} for audio pattern recognition. 
Different to the transformer-based models,
which operate on the raw audio signal,
this takes log-Mel spectrograms as input. 
\review{
CNN14 has $6$ convolutional blocks with two layers each, each followed by max pooling.
Convolution layers have a $3\times3$ kernel and a stride of $1\times1$, whereas max pooling layers use a stride of $2\times2$.
After the last convolution layer, features are pooled using both mean and max pooling, and subsequently fed into two linear layers.
Dropout with a probability of 0.2 is applied after every each convolution block.
Log-Mel spectrograms are computed with 64 Mel bins, a window size of 32\,ms, and a hop size of 10\,ms.
}
Note that the {\cnn} model is not pre-trained,
\ie it is always trained from scratch in our experiments.
We train for $60$ epochs,
with a learning rate of $.01$,
and a batch size of $64$
using \ac{SGD} with a Nesterov momentum of $.9$.
We select the model that performs best on the validation set.


\subsection{Datasets}
\label{subsec:datasets}


We used the \textbf{\msppodcast} corpus~\citep{lotfian2019msppodcast} (v1.7)
to run multitask training on the three dimensions of arousal, dominance, and valence
for speech from podcast recordings.
\review{The original labels cover a range from $1$ to $7$, 
which we map into the interval of $0$ to $1$.}
Its \emph{train} split contains $62$ hours of recordings. 
In-domain results are reported on the \emph{test-1} split,
which contains $21$ hours of audio
provided by $12,902$ samples ($54$\% female / $46$\% male)
from $60$ speakers ($30$ female / $30$ male). 
The samples per speaker 
vary between $42$ and $912$. 

We report cross-domain results \textbf{\iemocap} (Interactive Emotional Dyadic Motion Capture) dataset~\cite{busso2008iemocap},
which contains $12$ hours of scripted and improvised dialogues by ten speakers ($5$ female / $5$ male).
\review{It provides the same dimensional labels as {\msppodcast},
but in a range of $1$ to $5$, 
which we map to the interval $0$ to $1$.
Since we use the dataset only during evaluation, 
we do not apply the usual speaker cross-validation, but treat the corpus as a whole.}
It includes $10,039$ samples ($49$\% female / $51$\% male). 

Finally, we report cross-corpus results for valence on the test set of the
Multimodal Opinion Sentiment Intensity (\textbf{\mosi})~\citep{zadeh2016mosi} corpus.
The dataset is a collection of YouTube movie review videos
spoken by $41$ female and $48$ male speakers.
\review{They are annotated for sentiment on a $7$-point Likert scale ranging from $-3$ to $3$, 
which we map to the interval $0$ to $1$.}
The test set contains $1$ hour audio recordings
given as 685 samples ($51$\% female / $49$\% male),
annotated for sentiment.
\review{As the gender labels are not part of the distributed database,
we re-annotated them ourselves~\citep{wierstorf2023zenodo}.}

While sentiment is a different concept than valence,
as the former corresponds to an attitude held towards a specific object
and the latter more generally characterises a person's feeling~\citep{munezero2014they},
there is evidence that sentiment annotations can be decomposed to two constituents:
intensity and polarity~\citep{tian2018polarity},
which roughly correspond to arousal and valence.
We therefore expect some correlation between (predicted) valence
and (annotated) sentiment scores.


\subsection{Evaluation}
\label{subsec:evaluation}

Machine learning models for speech emotion recognition
are expected to work under different acoustic conditions and for different speakers.
To cover this,
we evaluate them for correctness, robustness, and fairness~\citep{zhang2019testing}.

\textbf{Correctness}
measures how well predictions match the ground truth.
The \acf{CCC} provides an estimate of how well the 
predicted distribution matches the ground truth one~\citep{lin1989ccc},
and is the typical measure for evaluating dimensional \ac{SER} models~\citep{ringeval2018avec}.

\textbf{Robustness} (\cf \cref{subsec:robustness})
measures how
\review{
model performance is affected by changes
to the input signals
such as adding background noise.
}
Applying changes to the input signals 
must be carefully done for \ac{SER},
as they might affect the ground truth label~\citep{jaiswal2021robustness,parada2017augmentation}.
We focus on testing the \review{robustness} of the models
against
\review{
data augmentations
that 
do not change the human perception
of the underlying emotion.
We select the following five augmentations 
from \citet{jaiswal2021robustness}
to enable direct comparison with previous results:
\emph{Natural soundscapes} adds a randomly selected sample
from the natural class of the ESC-50 dataset~\citep{piczak2015dataset}
with a \ac{SNR} of 0\,dB, 10\,dB or 20\,dB;
\emph{Human, non-speech} adds a randomly selected sample
from the human class of the ESC-50 dataset
with a \ac{SNR} of 0\,dB, 10\,dB or 20\,dB;
\emph{Interior/domestic} adds a randomly selected sample
from the interior class of the ESC-50 dataset
with a \ac{SNR} of 0\,dB, 10\,dB or 20\,dB;
\emph{Speed up segment} selects a random segment of $10$\% to $20$\% length
within the utterance
and increases its speed by $1.25$;
\emph{Fade-in/fade-out} decreases or increases the amplitude
of the signal by 2\% every second.
}

\textbf{Fairness} (\cf \cref{subsec:fairness})
evaluates if the model predictions show biases
for certain protected attributes like race, gender, or age~\citep{corbett-davies2018fairness}.
We focus on gender
due to the lack of sufficient available information and/or datasets for other attributes.
For regression problems,
there is no clear definition how to measure fairness,
but most approaches try to achieve an equal average
expected outcome for population A and B~\citep{fitzsimons2019fairness}.
We measure fairness by estimating the gender fairness score as the difference in the correctness metric (CCC) between female and male groups.
A positive gender fairness score indicates a better performance of the model for female speakers.


\section{Evaluation}
\label{sec:evaluation}

We begin our investigation with a thorough evaluation of transformer-based models.
\review{We show that valence is the primary beneficiary of pre-training as it enables the models to implicitly learn linguistic information during the fine-tuning of the transformer layers.}
Utilising a comprehensive testing scheme,
we attempt to identify how different aspects of foundation models
impact performance and generalisation.
We place particular emphasis on \emph{robustness} and \emph{fairness},
which are critical considerations for \ac{SER} systems targeted to real-world applications.


\subsection{Can foundation models close the performance gap for valence?}
\label{subsec:valence_gap}

\emph{\textbf{Answer}}:
\review{The best models achieve a similar performance for arousal and dominance 
as non-transformer architectures \citep{li2021contrastive},
but improve the \ac{CCC} score for valence by $.26$
and close the performance gap for valence.}

\begin{figure}[t]
    \centering
    \includegraphics[width=\columnwidth]{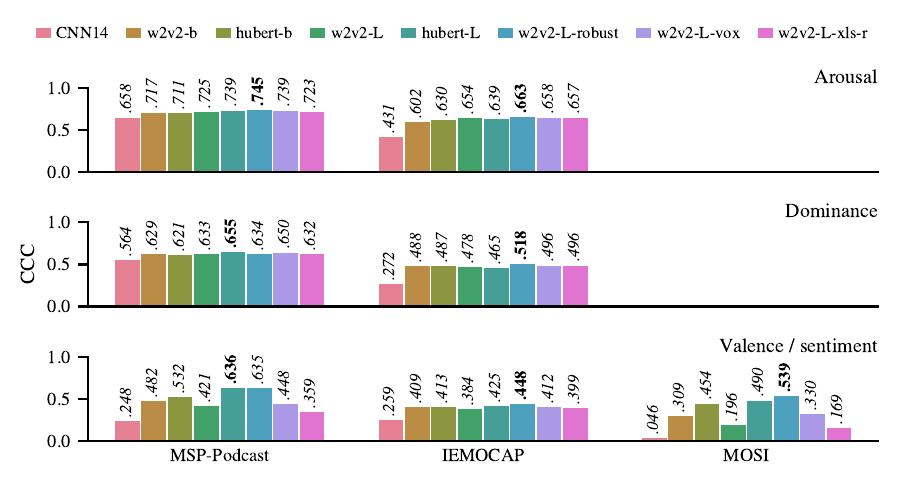}
    \caption{
        \ac{CCC} scores for arousal, dominance, valence ({\msppodcast} / {\iemocap}), and sentiment ({\mosi}).
        All models have been trained for emotional dimension prediction using multitasking \review{only} on {\msppodcast},
        and subsequently evaluated on its test set (in-domain),
        as well as to the test set of {\mosi}
        and the entire {\iemocap} dataset (cross-corpus).
    }
    \label{fig:results}
\end{figure}

\noindent
\emph{\textbf{Details:}}
In \cref{fig:results},
we show in-domain and cross-domain \ac{CCC} performance for different {\wtov} and {\hubert} models 
as well as for the {\cnn} baseline.

\review{
We first focus on arousal and dominance.
For {\msppodcast} (in-domain)
and {\iemocap} (cross-domain),
all transformer-based models score
very similar,
with {\wrobust} showing the overall best performance
by reaching a \ac{CCC} score of $.745$/$.634$ (arousal/dominance)
on {\msppodcast},
and $.663$/$.518$
on {\iemocap}.
For {\msppodcast}, results are similar
compared to the \ac{CCC} scores of $.745$/$.655$ 
achieved by \citet{li2021contrastive}
and $.757$/$.671$
achieved by \citet{srinivasan2021representation}.
}

\review{
For valence,
we see a larger fluctuation of \ac{CCC} scores
for different transformer models
ranging from $.359$ for {\wxlsr}
to $.636$ for {\hbase},
both on {\msppodcast}.
Overall,
{\wrobust} shows again the best overall performance
by reaching a \ac{CCC} score of $.635$ on {\msppodcast},
$.448$ on {\iemocap},
and $.539$ for predicting sentiment on {\mosi}.
For {\msppodcast}, results are better
compared to the \ac{CCC} score of $.377$
achieved by \citet{li2021contrastive}
and similar to $.627$
by \citet{srinivasan2021representation}
achieved with a model distilled from an audio + text based teacher.
}


\subsection{Does explicit linguistic information further improve performance?}
\label{subsec:fusion}

\emph{\textbf{Answer}}:
Adding linguistic information does not improve predictions for arousal and dominance,
and only in some cases for valence.
However, especially models pre-trained on multiple languages seem to benefit when tested on English speech.

\begin{figure}[t]
    \centering
    \includegraphics[width=\columnwidth]{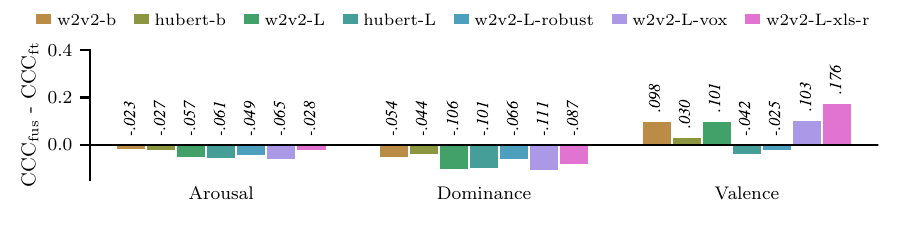}
    \caption{
        Text \& audio fusion results for arousal, dominance, and valence prediction on {\msppodcast}.
        Embeddings from the already fine-tuned models are concatenated with BERT embeddings extracted from automatic transcriptions, whereupon a two-layer feed-forward neural network is trained. 
        We show the difference to results with the fine-tuned (ft) models from \cref{fig:results}.
    }
    \label{fig:fusion}
\end{figure}

\noindent
\emph{\textbf{Details:}}
To evaluate whether adding linguistic information improves the predictions, the following experiment is conducted: a regression head is pre-trained, using as input pooled BERT embeddings in addition to the pooled states of the fine-tuned transformer models.

BERT (Bidirectional Encoder Representations from Transformers) is a transformer model for natural language, pre-trained on English language corpora consisting of more than $3$ billion words~\cite{devlin2019bert}. 
The BERT embeddings have a dimensionality of 768 and are extracted from the transcriptions generated by the \emph{wav2vec2-base-960h} speech recognition model\footnote{\url{https://huggingface.co/facebook/wav2vec2-base-960h}}.
The fusion is done by concatenating the representations of both modalities.
As regression head, exactly the same architecture as for the fine-tuning of {\wtov} and {\hubert} models is employed.
For training, the weights of both models are frozen. 
The training is done with multi-target \ac{CCC}-loss for a maximum of 100 epochs, with early stopping based on \ac{CCC} development set performance. 

In \cref{fig:fusion}, we report deviations from the results achieved with the fine-tuned acoustic models alone (\cf \cref{fig:results}).
We can see that a fusion with embeddings from the text domain helps with valence, but not with arousal and dominance, where performance actually deteriorates.
This is in line with our previous findings, where we also found that introducing linguistic information sometimes hampered performance for those two dimensions on {\msppodcast}~\citep{triantafyllopoulos2021multistage}.
What is interesting, though, are the relatively large differences between the models, and that, especially, our best models {\hlarge} and {\wrobust} do not improve.
The models that benefit most are the two multi-lingual models {\wvox} and {\wxlsr}, showing that models pre-trained on multiple languages gain from a fusion with text features from the test set domain language.


\subsection{Do the models implicitly learn linguistic information?}
\label{subsec:tts}

\emph{\textbf{Answer}}:
The models implicitly capture linguistic information from the audio signal.
The extent in which they learn sentiment during fine-tuning
depends on the data used for pre-training
(\eg multi-lingual data makes it more difficult).
Generally, we see that valence performance correlates with a model's ability to predict sentiment.

\begin{figure}[t]
    \centering
    \includegraphics[width=\columnwidth]{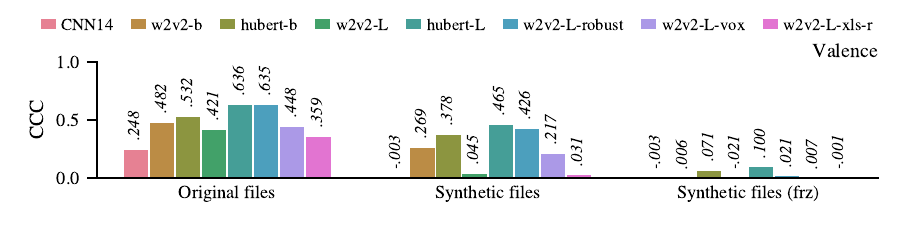}
    \caption{
        \ac{CCC} performance for valence on the original and synthetic files on {\msppodcast}.
        We see that models with a high performance on the original files are more sensitive to sentiment (\cf left and center section).
        To prove that a fine-tuning of the transformer layers is required to learn linguistic content, we additionally show the correlation for models where the transformer layers were frozen (frz) during training (\cf \cref{subsec:frozen}).
    }
    \label{fig:tts-corr}
\end{figure}

\noindent
\emph{\textbf{Details:}}
Previous findings suggest that during fine-tuning,
the models implicitly learn linguistic information.
To asses how sensitive the models are to linguistic content,
we generated a synthesised version of a subset of the test set
from the transcriptions of {\msppodcast}.\footnote{
Partial audio transcripts are available with {\msppodcast} v1.9
and cover 55\% of the \emph{test-1} split from v1.7 we used for our experiments.
}
In \cref{fig:tts-corr}, we finally show \ac{CCC} performance for valence
on the original and synthesised files for all models.
We see that performance gaps between the models in \cref{fig:results}
are directly linked with their ability to predict sentiment.
Models reaching a high performance on the original files also do so on their synthetic versions and vice versa.
However, to learn linguistic content, a fine-tuning of the transformer layers is essential.
If we predict the synthetic test set with models
where the transformer layers were frozen during training (\cf \cref{subsec:frozen}),
correlation drops to almost zero.

\review{
This finding
is also important
for works doing in-domain training
on IEMOCAP,
as parts of the conversations
are scripted
which results in a leakage of text information
that may result in overoptimistic results~\citep{pepino2020fusion}
when that text information is exploited by transformer models.
Furthermore,
our models may inherit
similar biases
as those found in NLP models~\citep{triantafyllopoulos2022probing}.
}


\subsection{How important is a fine-tuning of the transformer layers?}
\label{subsec:frozen}

\emph{\textbf{Answer}}:
Fine-tuning the transformer layers is necessary to obtain state-of-the-art performance,
in particular for the valence dimension.
The highest gain is observed for {\hlarge} and {\wrobust},
which are the models that benefit the least from a fusion with text.

\begin{figure}[t]
    \centering
    \includegraphics[width=\columnwidth]{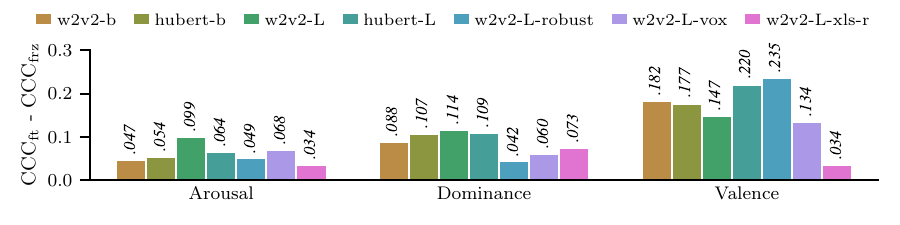}
    \caption{
        Difference of fine-tuned (ft) to frozen (frz) \ac{CCC} performance for arousal, dominance, and valence prediction on {\msppodcast}.
        The fine-tuned results are from \cref{fig:results}, where transformer and output layers are jointly trained.
        For the frozen results, we keep all transformer layers frozen and simply train the output head.
        Results show that fine-tuning the transformer layer is worth the computational cost it incurs.
    }
    \label{fig:wo-ft}
\end{figure}

\noindent
\emph{\textbf{Details:}}
So far, we have fine-tuned all transformer layers along with the added output layer.
However, 
practitioners often choose to use a pre-trained model as a frozen feature extractor,
and subsequently train just an output layer on the generated embeddings.
Nevertheless, prior studies have shown
that it is necessary to fine-tune several or all layers on the target task
to get good downstream performance\,\citep{kong2019panns, wang2021finetuned, liu2020mockingjay, triantafyllopoulos2021role}.
In this sub-section,
we experiment with training only the last output layer and keeping all others frozen.
This is compared to our previous experiments
where we jointly fine-tune the last layer and the transformer layers.

\cref{fig:wo-ft} shows the difference between \ac{CCC} values
for the fine-tuned and frozen models.
We observe performance gains when fine-tuning in all cases,
demonstrating that fine-tuning of the transformer layers is necessary.
Moreover, the models that see the biggest performance gain 
are {\hlarge} and {\wrobust}.
In \cref{subsec:fusion},
these models were found to benefit less from additional text information.
These findings indicate
that a fine-tuning of the transformer layers
enables the models to capture the linguistic information
needed to perform well on valence.


\subsection{Do the models generalise better across different domains?}
\label{subsec:generalisation}

\emph{\textbf{Answer}}:
Transformer-based models generalise better than a non-transformer baseline.

\noindent
\emph{\textbf{Details:}}
As we see a similar trend for different transformer-based models
between in-domain and cross-corpus results in \cref{fig:results},
we focus on the best-performing one ({\wrobust}).
The drop in \ac{CCC} between in-domain and cross-corpus results for {\wrobust} on {\iemocap} is $11$\% for arousal,
$21$\% for dominance,
and $30$\% for valence on IEMOCAP, and $15$\% for sentiment on {\mosi}.
For {\cnn}, the drop in \ac{CCC} is $34$\% for arousal,
and $52$\% for dominance, while for valence,
we do not estimate the drop in cross-domain performance
as the in-domain \ac{CCC} is already very low.
The drop in \ac{CCC} is smaller for {\wrobust} for arousal and dominance,
indicating that transformer-based models generalise better.
For valence, we cannot derive a final conclusion,
but the trend we see for sentiment in {\mosi} seems very promising.


\subsection{Does more data during pre-training lead to better performance?}
\label{subsec:influence_data}

\emph{\textbf{Answer}}:
For arousal and dominance,
all tested models perform equally well,
whereas with respect to valence / sentiment the data used for pre-training has a strong effect.
Mixing data from several domains leads to a considerable improvement for {\wrobust}
compared to {\wlarge},
which is only trained on clean speech.
However, {\hlarge}, 
which uses the same pre-training data as {\wlarge},
which is also trained on clean speech,
still performs as good as {\wrobust}.
For models pre-trained on multi-lingual data,
we see a performance drop when tested on English speech.

\noindent
\emph{\textbf{Details:}}
To understand
what influence the size and domain of the pre-training data
have on downstream performance,
we included several {\wtov} models
with same large architecture but different pre-training (see \cref{tab:models}). 

The results in \cref{fig:results} show only differences
in terms of \ac{CCC} between the transformer models
for valence and sentiment,
not for arousal or dominance.
\review{Previous studies uniformly report 
that {\hubert} outperforms {\wtov}
which is replicated by our results 
with {\wbase} showing a smaller CCC than {\hbase}
for the valence task on {\msppodcast} and {\iemocap}, 
and for the sentiment task on {\mosi}. 
The increase in performance 
for {\wrobust} 
compared to {\hlarge} 
is most likely explained 
by the additional 3k hours of telephone conversations 
used for pre-training.}
However, by comparing {\wvox} and {\wxlsr},
it also becomes clear
that more data does not necessarily lead to better results. 
Though both models are trained on significantly more data
than {\hlarge} and {\wrobust} ($100$k and $463$k vs $63$k hours),
they perform clearly worse.
Notably, both were pre-trained on multiple languages. 
Since the databases we use for evaluation contain only English speakers,
this could be a disadvantage to models that are exclusively pre-trained on English.


\subsection{Does a larger architecture lead to better performance?}
\label{subsec:larger_architecture}

\emph{\textbf{Answer}}:
A larger architecture does not lead to better performance per se.
Larger architectures using different data during pre-training might perform worse than smaller architectures. 

\noindent
\emph{\textbf{Details:}}
\review{We can draw some conclusions 
about the influence 
that the size of the architecture
has on performance
based on \cref{fig:results}.}
The size of the architecture, \ie base vs large,
seems not to be the decisive point: 
\review{{\wbase} and {\wlarge} show very similar performance.}
\review{In addition,}
the small models {\wbase} and {\hbase} 
have comparable performance 
to the large models {\wlarge}, {\wvox}, and {\wxlsr} 
for arousal and dominance,
both in- and cross-domain. 
For valence, the small models outperform {\wlarge}, {\wvox}, and {\wxlsr}
in most cases for {\msppodcast} and {\mosi},
and achieve a similar performance on {\iemocap}.

\subsection{Are the models robust against changes to the input signals?}
\label{subsec:robustness}

\emph{\textbf{Answer}}:
\review{
The tested models are reasonably robust against changes to the input signals,
with {\wrobust} showing the highest and {\hbase} the lowest robustness.
}

\begin{figure}[t]
    \centering
    \includegraphics[width=\columnwidth]{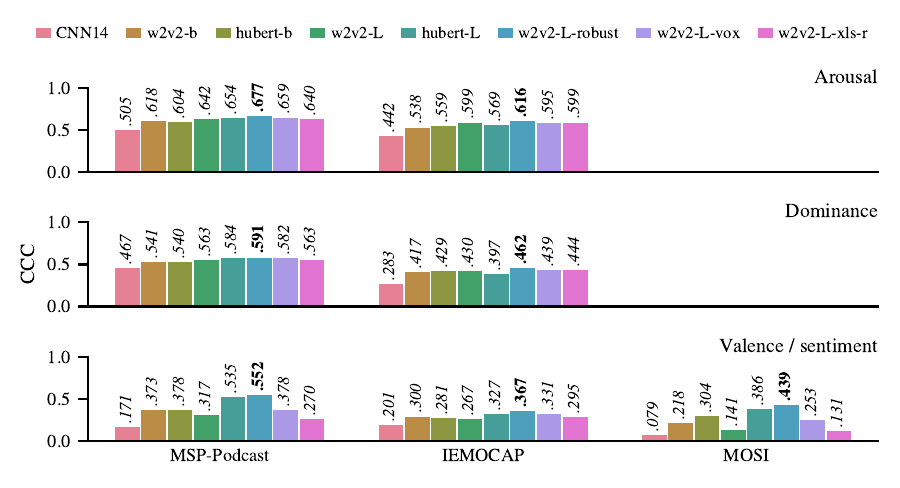}
    \caption{
        \review{
        \ac{CCC} scores for arousal, dominance,
        valence ({\msppodcast}/{\iemocap}), and sentiment ({\mosi})
        when augmenting the test data.
        The scores are averaged over all eleven different
        augmented versions of the test data.
        }
    }
    \label{fig:robustness}
\end{figure}

\noindent
\emph{\textbf{Details:}}
\review{
\cref{fig:robustness} summarises the average \ac{CCC} scores of the models
averaged over all augmentations described in \cref{subsec:robustness}.
All models show a drop in \ac{CCC}
compared to the \ac{CCC} scores for the clean data
from \cref{fig:results}.
{\wrobust} has now the highest \ac{CCC} score
for all datasets and all dimensions.
The average change in \ac{CCC} for {\wrobust} is $-0.068$.
The model with the highest average change in \ac{CCC}
is {\hbase} ($-0.108$).
The model with the lowest average change in \ac{CCC}
is {\cnn} ($-0.047$),
which is mostly due to its results for {\iemocap}
for which it shows no impairment
of its relatively low performance
by augmentations.
}

\begin{table*}[!t]
    \centering
    \caption{
        \review{
        Change in \ac{CCC} for {\wrobust} predictions
        on augmented data
        compared to its predictions on clean data.
        }
    }
    \begin{threeparttable}
        \begin{color}{black}
        \resizebox{\columnwidth}{!}{
        \begin{tabular}{llrrrrrrr}
            \toprule
             Augmentation & SNR & \multicolumn{2}{c}{Arousal} & \multicolumn{2}{c}{Dominance} & \multicolumn{3}{c}{Valence / sentiment} \\
             &  & MSP-Podcast & IEMOCAP & MSP-Podcast & IEMOCAP &         MSP-Podcast & IEMOCAP &   MOSI \\
            \midrule
            Natural soundscapes & 0 dB  & -.145 & -.109 & -.081 & -.119 & -.221 & -.183 & -.210 \\
                                & 10 dB & -.038 & -.018 & +.001 & -.035 & -.051 & -.064 & -.071 \\
                                & 20 dB & -.019 & -.009 & +.012 & -.018 & -.003 & -.015 & -.025 \\
            Human, non-speech   & 0 dB  & -.254 & -.205 & -.195 & -.167 & -.278 & -.214 & -.264 \\
                                & 10 dB & -.053 & -.036 & -.036 & -.058 & -.080 & -.092 & -.143 \\
                                & 20 dB & -.011 & -.001 & +.005 & -.017 & -.007 & -.030 & -.036 \\
            Interior/domestic   & 0 dB  & -.124 & -.089 & -.081 & -.097 & -.170 & -.160 & -.149 \\
                                & 10 dB & -.026 & -.011 & -.003 & -.029 & -.031 & -.049 & -.059 \\
                                & 20 dB & -.011 & -.004 & +.007 & -.014 & -.005 & -.015 & -.030 \\
            Speed up segment    &       & -.080 & -.030 & -.099 & -.056 & -.062 & -.069 & -.113 \\
            Fade-in/fade-out    &       & -.001 & -.001 &  .000 & -.001 &  .000 &  .000 & -.001 \\
            \bottomrule
        \end{tabular}
        }
        \end{color}
    \end{threeparttable}
    \label{tab:robustness}
\end{table*}

\review{
\cref{tab:robustness} shows changes in \ac{CCC} for single augmentations
for each dataset and dimension
for the best performing model {\wrobust}.
The performance of the model is only sligthly affected
(absolute change in \ac{CCC} score below $.05$)
for added background sounds with a \ac{SNR} of $20$\,dB
or a fade-in/fade-out of the signal.
When speeding up parts of the signal or adding background sounds
with more severe \acp{SNR}
the change in \ac{CCC} can be up to $-.278$.
The model investigated on the same augmentations
by \citet{jaiswal2021robustness}
shows an equal drop in \ac{UAR}
when adding background sounds
with $0$\,dB, $10$\,dB, $20$\,dB \ac{SNR}
of at least $-.30$.
{\wrobust} is more robust when adding
background sounds with a moderate \ac{SNR}.
It shows a drop in \ac{CCC} of up to to $-.28$ for $0$\,dB SNR,
but only a drop in \ac{CCC} of up to $-.036$ for $20$\,dB SNR.
Whereas the model investigated by \citet{jaiswal2021robustness}
is similar affected by adding
\emph{human, non-speech}, \emph{interior/domestic}, or \emph{natural sounds}
as background sounds,
{\wrobust} is the most affected when adding \emph{human, non-speech} sounds (average drop in \ac{CCC} of $-.103$),
and the least when adding \emph{interior/domestic} sounds
(average drop in \ac{CCC} of $-.055$).
}


\subsection{Are the models fair regarding the gender of the speaker?}
\label{subsec:fairness}

\emph{\textbf{Answer}}:
Models are \review{more} fair for arousal and dominance
\review{than for valence}.
For valence,
most models show a higher \ac{CCC} for females than for males.

\begin{figure}[t]
    \centering
    \includegraphics[width=\columnwidth]{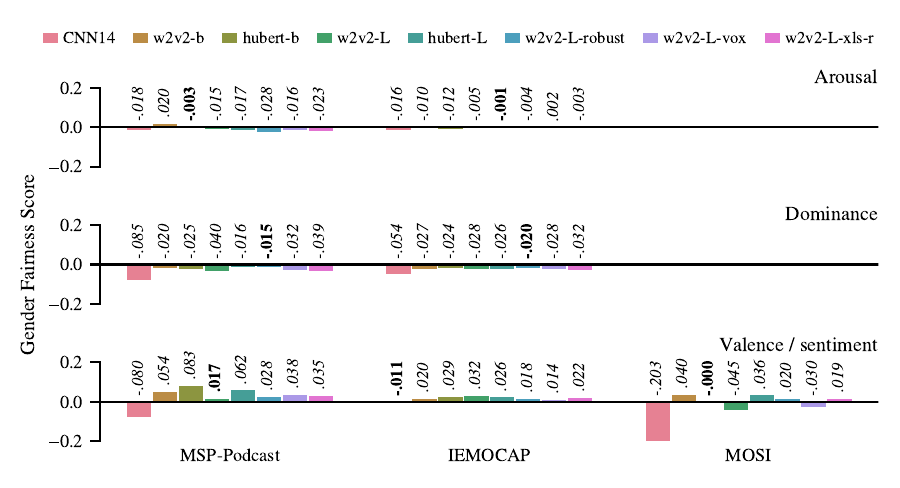}
    \caption{Gender fairness scores for arousal, dominance, valence ({\msppodcast} / {\iemocap}), and sentiment ({\mosi}). The gender fairness score is given by $\text{CCC}_\text{female} - \text{CCC}_\text{male}$. A positive value indicates that the model under test performs better for female speaker and a negative value that it performs better for male speaker. A model with desired equal performance would have a gender fairness score of 0.
    }
    \label{fig:fairness}
\end{figure}

\noindent
\emph{\textbf{Details:}}
\cref{fig:fairness} shows gender fairness scores
for the speakers in {\msppodcast}, {\iemocap}, and {\mosi}.
As introduced in \cref{subsec:evaluation},
the gender fairness score is expressed
by the difference in \ac{CCC} between female and male speakers
with positive values indicating higher values of the underlying metric for females.
For {\msppodcast},
nearly all models show a slightly worse female \ac{CCC} for arousal and dominance. 
For {\iemocap},
nearly all models show a slightly better female \ac{CCC} for arousal and dominance.

For valence in {\msppodcast} and {\iemocap}, 
most models show a better \ac{CCC} for female speakers than male ones -- 
with the exception of {\cnn}. 
For sentiment in {\mosi}, the {\cnn} model shows a bias towards better performance for male speaker,
whereas all other models show very small biases in the different direction.

Averaging over all databases and dimensions the model with the best gender fairness score is {\wlarge} with $.007$, followed by {\wvox} with $.015$, {\wxlsr} with $.018$, {\wrobust}, with $.019$, {\hbase} with $.025$, {\hlarge} with $.027$, and {\wbase} with $.029$ up to {\cnn} with $-.043$.


\subsection{Is performance equal across all speakers?}
\label{subsec:speakers}

\emph{\textbf{Answer}}:
Performance for the best foundation models is similar between most speakers in {\msppodcast},
but can deteriorate to low \ac{CCC} values for some speakers.

\begin{figure}[t]
    \centering
    \includegraphics[width=\columnwidth]{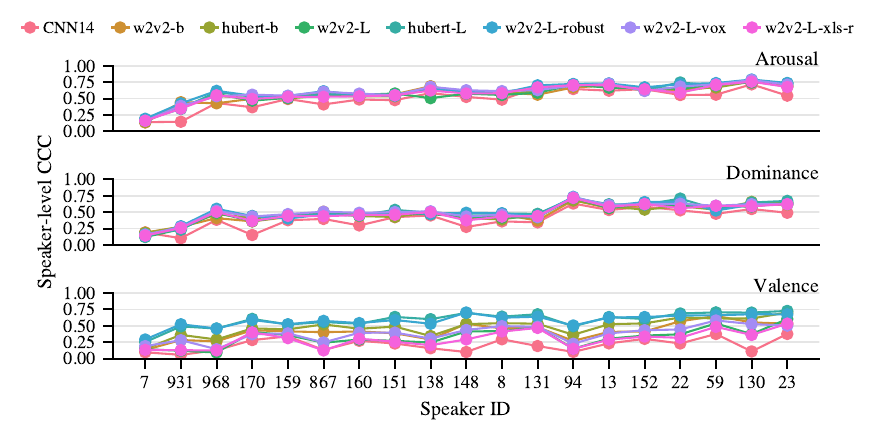}
    \caption{
        Speaker-level performance (\acs{CCC}) on {\msppodcast} for the different models.
        We only use speakers with at least 200 test set samples for robust \acs{CCC} estimates.
        All models show low \ac{CCC} for at least one speaker on all 3 tasks.
        Speakers have been ordered according to the mean \acs{CCC} over all dimensions and models.
    }
    \label{fig:speaker:plots}
\end{figure}

\noindent
\emph{\textbf{Details:}}
The performance of speech processing is dependent on individual speaker characteristics~\citep{doddington1998sheep}.
This has led several prior \ac{SER} works to target personalisation to different speakers~\citep{rudovic2018personalized, triantafyllopoulos2021deep, sridhar2022unsupervised}.
To investigate this phenomenon for transformer-based models, we examine the \emph{per-speaker performance}, where instead of computing a global \ac{CCC} value over all test set values, we compute one for each speaker.
As discussed (\cf \cref{subsec:datasets}), the {\msppodcast} test set consists of 12902 samples from 60 speakers; however, the samples are not equally distributed across them (minimum samples: 41, maximum samples 912).
In order to make our subsequent analysis more robust, we only keep speakers with more than 200 samples, resulting in 19 speakers.
We use bootstrapping, where we randomly sample (with replacement) 200 samples from each speaker to compute the \ac{CCC}.
This process is repeated 1000 times, and we report the mean value.

Our results are presented in \cref{fig:speaker:plots}.
For visualisation purposes, we ordered speakers based on the average \ac{CCC} value over all models and across arousal, dominance, and valence.
For arousal and dominance models perform well for most speakers,
and show similar performance.
For speakers 7 and 931 all models show a low CCC, whereas for speaker 931 the {\cnn} model performs worse than the others.
For valence, CCC values per speaker differ between models
replicating the findings of \cref{fig:results}.
The best model ({\wrobust}) performs relatively similar for most of the speaker groups and shows only a drop for speaker 7, a similar result as for valence and dominance.

Different models broadly, but not perfectly, agree on `good' and `bad' speakers, with pairwise Spearman correlations ranging from $.960$ to $.725$ for arousal, $.972$ to $.825$ for dominance, and $.947$ to $.333$ for valence.
This could be a manifestation of the underspecification phenomenon plaguing machine learning architectures~\citep{d2020underspecification}, as models which have similar performance on the entire test set, nevertheless behave differently across different subsets of it.


\subsection{Why do foundation models generalise so well?}
\label{subsec:tsne}

\emph{\textbf{Answer}}:
Even without pre-training,
the latent space provided by the transformer architecture generalises better than {\cnn},
as it abstracts away domain and speaker.
Pre-training marginally improves arousal and dominance performance
but is critical for valence.

\begin{figure}[t]
    \centering
    \includegraphics[width=\columnwidth]{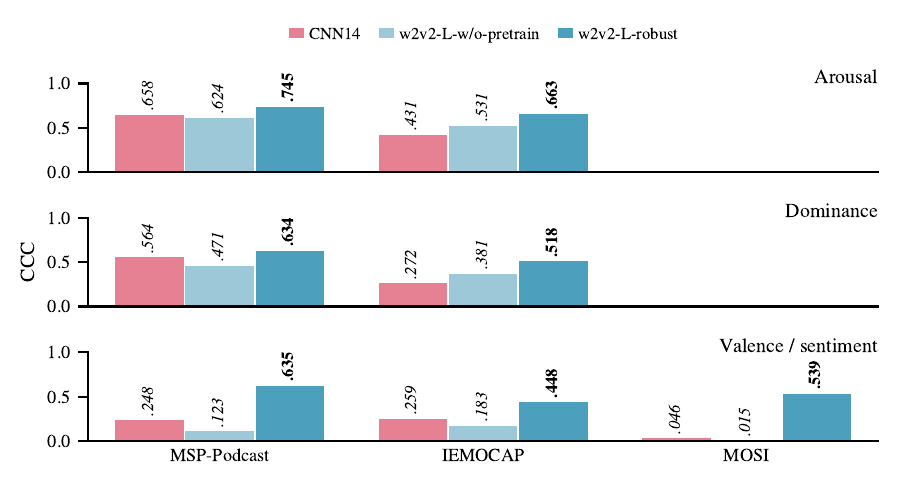}
    \caption{
        \ac{CCC} performance of randomly-initialised {\wtov} model (\emph{\wwopre}) on in-domain and cross-corpus arousal, dominance, valence / sentiment prediction.
        We compare the performance with that of \emph{\cnn} and \emph{\wrobust}. 
        We observe that valence and sentiment benefit massively from pre-training, without which {\wtov} performs worse than a classic CNN approach.
    }
    \label{fig:wo-pretrain}
\end{figure}

\noindent
\emph{\textbf{Details:}}
So far,
we were able to confirm the superiority of transformer-based models.
However, even though pre-training seems important,
it remains unclear to what extent the transformer architecture itself contributes to that success. 
To shed more light into this, we trained {\wtov} from a random initialisation. 
As our architecture,
we chose the large {\wtov} architecture,
which is also used by the best performing model {\wrobust}. 
In the following, we will refer to this model as {\wwopre}. 

We trained the model for $50$ epochs
and selected the best checkpoint
according to the performance on the development set (epoch 17).\footnote{
Even though we used the same data ({\msppodcast}) for fine-tuning,
we expected it would take longer for the model to convert if we start from scratch.
Also, this time we trained all encoder layers (including the CNN ones). 
Apart from that we followed the methodology described in \cref{subsec:architecture}.
}
In \cref{fig:wo-pretrain},
we compare in- and cross-domain performance with {\cnn} and {\wrobust}. 
We see that especially valence / sentiment detection benefits massively from pre-training
(both in-domain and cross-domain),
and that without pre-training {\wtov} performs in most cases worse than {\cnn}.

\begin{figure}[t]
    \centering
    \includegraphics[width=\columnwidth]{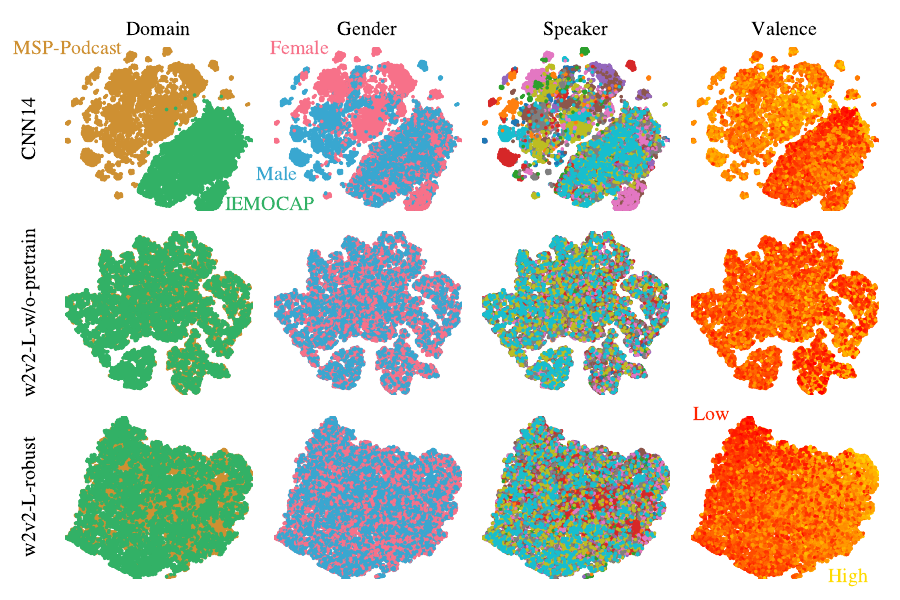}
    \caption{
        Visualisation of embeddings extracted with different models overlayed with meta information for a combined dataset of {\msppodcast} and {\iemocap}. 
        We observe that the latent space of {\wtov} offers a better abstraction from domain, gender,  and speaker compared to the {\cnn} baseline -- even without pre-training. 
        However, only a pre-trained model is able to separate low from high valence. 
        To reduce the dimensionality of the latent space, we applied T-SNE~\citep{maaten2008tsne}.
    }
    \label{fig:embeddings}
\end{figure}

In the introduction of {\wtov}, \citet{baevski2020wav2vec} postulate
that  pre-training helps learn more general representations
that abstract away from speaker or background information.
However, it is not entirely clear
if these benefits are a result of pre-training
or are a consequence of the specific inductive biases introduced by the architecture.
To investigate this,
we compare embeddings extracted with {\cnn}, {\wwopre}, and {\wrobust},\footnote{
We use average pooling on the output of the last CNN layer for {\cnn}
and the last transformer layer for {\wtov}.
}
which are shown in \cref{fig:embeddings}.
The embeddings are projected to two dimensions using t-SNE~\citep{maaten2008tsne}
and different information is chromatically superimposed.

For {\cnn}, two main clusters almost perfectly separate the two data sources {\msppodcast} and {\iemocap},
whereas several smaller blobs represent gender groups and individual speakers.
In fact, speaker and domain are more pronounced than valence information. 
Hence, similar emotional content can translate into entirely different latent representations. 
In contrast,
the latent space of both {\wtov} models shows no clusters for domain, gender, or speaker.
The architecture itself seems to introduce specific inductive biases
which are well-suited to learning robust representations.
Nevertheless, only the pre-trained model ({\wrobust}) shows a smooth transition from low to high valence scores,
showing that pre-training is still necessary for good downstream performance.
Moreover, the strong speaker dependency presented in \cref{subsec:speakers}
shows that the two dimensional t-SNE visualisations help comparing generalisation abilities between models,
but are not necessarily sufficient for deriving conclusions \wrt generalisation over different factors.


\section{Efficiency}
\label{sec:efficiency}

For our last experimental evaluation,
we focus on efficiency.
We concentrate on three facets:
\emph{optimisation stability},
\emph{computational complexity},
and \emph{data efficiency}.


\subsection{Does pre-training help with training stability and convergence?}
\label{subsec:stability}

\emph{\textbf{Answer}}:
A pre-trained model reduces the number of epochs needed to converge
and improves performance stability across training runs with different seeds.

\begin{figure}[t]
    \centering
    \includegraphics[width=\columnwidth]{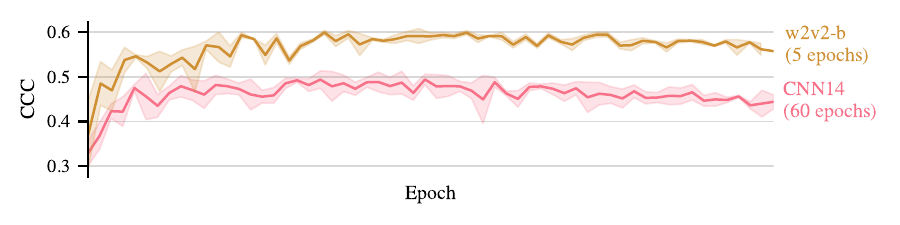}
    \caption{
        Mean and standard deviation of development set performance on {\msppodcast} across three training runs.
        Compared to \emph{\cnn}, \emph{\wbase} converges earlier and shows less fluctuation.
    }
    \label{fig:stability}
\end{figure}

\noindent
\emph{\textbf{Details:}}
To balance the effects of randomness
(either in the initialisation of network weights or the data sampling),
it is a common strategy to perform several runs with different random seeds. 
Starting from pre-trained weights, however, we expect less volatility~\citep{erhan2010does, neyshabur2020being}. 
\cref{fig:stability} shows the mean and standard deviation
over the performance on the development set across three trials for {\cnn} and {\wbase}. 
{\cnn} shows a constant jittering across all 60 epochs,
whereas {\wbase} converges faster
and we can reduce the number of epochs to $5$. 


\subsection{How many transformer layers do we really need?}
\label{subsec:prune}

\emph{\textbf{Answer}}:
We can reduce the number of transformer layers to 12 without a degradation in performance.
With less than 12 layers we begin to see a negative effect on valence.

\begin{figure}[t]
    \centering
    \includegraphics[width=1.\linewidth]{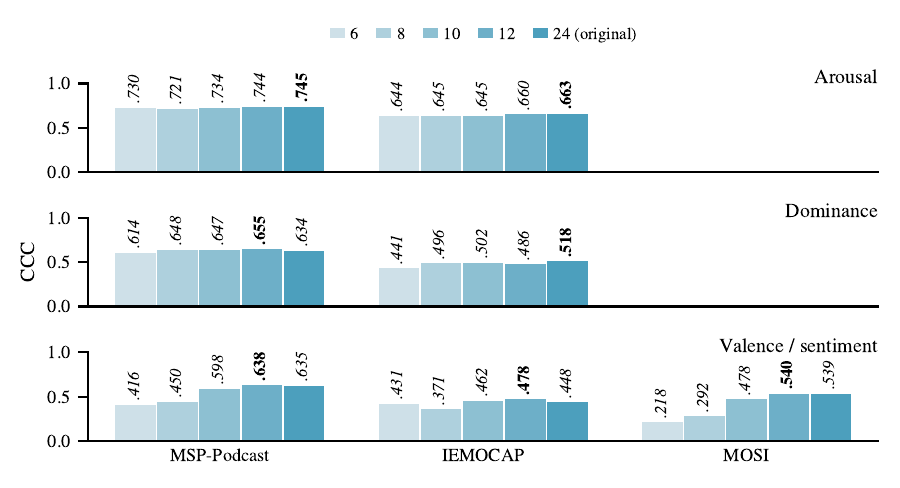}
    \caption{
        \ac{CCC} scores for arousal, dominance, and valence / sentiment for \emph{\wrobust}
        and pruned versions.
        The legend shows the number of bottom layers kept during fine-tuning.
        We see that half of the layers can be removed without any loss in performance.
    }
    \label{fig:pruning}
\end{figure}

\noindent
\emph{\textbf{Details:}}
In \cref{subsec:larger_architecture},
we mentioned that {\wbase} and {\hbase} outperform some of the large models.
From that, we concluded that the size of the architecture seems less important,
but it is rather the data used for pre-training that determines success. 
If this is really the case,
we should be able to partially reduce the size of a model without losing performance.

\citet{sajjad2021poor} investigated different layer pruning strategies
and identified top-layer dropping as the best strategy offering a good trade-off between accuracy and model size. 
Inspired by their findings,
we set up an experiment where we successively removed transformer layers from the top of the original pre-trained model before fine-tuning. 
In \cref{fig:pruning}, we report the effect on CCC for {\wrobust} (our overall best performing model). 
Results show that half of the layers can be removed without a loss in performance. 
We denote the resulting $12$-layer model as {\wrobustpruned}.
Only with 10 or less layers we actually begin to see a drop for valence / sentiment on {\iemocap} and {\mosi}. 
For arousal and dominance, we still achieve good performance with only $8$ layers.


\subsection{Can we reduce the training data without a loss in performance?}
\label{subsec:data_reduction}

\emph{\textbf{Answer}}:
A reduction of training samples without loss in performance is only possible for arousal and dominance.

\begin{figure}[t]
    \centering
    \includegraphics[width=1.\linewidth]{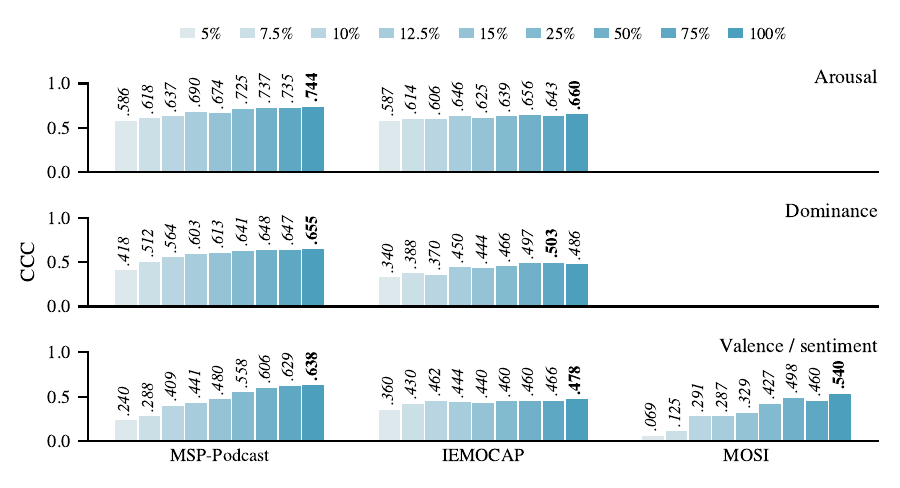}
    \caption{
        \ac{CCC} scores for arousal, dominance, and valence / sentiment for \emph{\wrobust} on sparse training data.
        The legend shows the fraction of data used for fine-tuning.
        Please note that steps are not linear.
    }
    \label{fig:sparse}
\end{figure}

\noindent
\emph{\textbf{Details:}}
Reducing the amount of training data offers another way to speed up model building. 
To find out what effect the removal of training samples has,
we conducted an experiment where we fine-tuned several versions of the same pre-trained model
with different fractions of the training set ({\msppodcast}). 
We leave development and test set untouched.

\cref{fig:sparse} shows \ac{CCC}
for arousal, dominance, valence / sentiment on {\msppodcast}, {\iemocap} and {\mosi}.
For efficiency, we start from the reduced $12$-layer architecture and therefore compare results to {\wrobustpruned} (\cf \cref{subsec:prune}).
There is no noteworthy degradation for arousal and dominance when keeping close to the entire training set.
The only exception is dominance on {\iemocap}, where we achieve best results with just $75$\% of the data. 
For these dimensions, however, performance already saturates at $25$\% yielding a loss of less than $.02$ on {\msppodcast},
whereas for {\iemocap}, even $12.5$\% of the training samples seem sufficient to stay within a margin of $.05$.

Once again, it is a different story for valence. 
For {\msppodcast}, we see a constant improvement that only begins to decrease when reaching $75$\% of the data. For {\mosi}, we even see a boost in \ac{CCC} of almost $.1$ for the remaining $25$\%. 
However, in light of our findings from \cref{subsec:tts}, 
this does not come as a surprise. 
Providing more linguistic diversity makes it more likely a model can detect associations between key words and emotional context. 
What is a surprise, though, is that on {\iemocap}, using just $7.5$\% of the data, results in a drop of less than $.05$. 
A possible explanation is that the vocabulary of {\iemocap} does not resemble that of {\msppodcast} and that, therefore, the impact of linguistic information is limited. 
This would also explain why the differences in valence performance are less pronounced for {\iemocap} (\cf \cref{fig:results}).


\section{Summary}
\label{sec:summary}

We explored the use of (pre-trained) transformer-based architectures for speech emotion recognition.
In the previous sections, we dealt with several questions in isolation.
We now attempt a unified summary by collectively considering all findings.

\textbf{Effect of pre-training:}
pre-training is essential to get good performance (\cref{subsec:influence_data}),
especially for the valence dimension.
This is particularly evident
when training {\wtov} from a random initialisation (\cref{subsec:tsne}):
the model performs substantially worse on all three dimensions,
and its embeddings are unable to capture valence information.
In addition,
pre-training serves as a form of regularisation
which helps stabilise the training (\cref{subsec:stability}),
thus resulting in models which require less iterations,
and less data to train on (\cref{subsec:data_reduction}).
However, we were unable to determine a clear relationship of the form
`more pre-training data leads to better performance'.
In fact, downstream performance can be negatively impacted by the introduction of more data,
as seen by the comparison between {\wvox} and {\wxlsr},
which differ only in the fact that {\wxlsr} has been trained on more (and more diverse) data,
yet performs worse on all three dimensions.

\textbf{Generalisation:}
transformer-based models show very good cross-corpus generalisation (\cref{subsec:influence_data}),
\review{robustness} (\cref{subsec:robustness}),
and appear invariant to domain, speaker, and gender characteristics (\cref{subsec:tsne}).
These are all very important traits for any model
that is intended for production use in realistic environments.
However, they seem to stem primarily from the architecture
rather than the pre-training
as they are also evident in models initialised from random weights (\cref{subsec:tsne}).
We also showed that several self-attention layers can be removed
without hampering downstream performance (\cref{subsec:prune}),
though they might still be necessary for successful pre-training.

\textbf{Fairness:}
fairness remains a challenging topic for contemporary machine learning architectures.
Community discussions primarily concern the issue of \emph{group fairness}.
In the present, we investigate this for the only group variable available in our datasets:
gender (\cref{subsec:fairness}),
where we observe that transformer-based architectures are more fair than the {\cnn} baseline.
However, we argue that \emph{individual fairness} is important for \ac{SER}.
This refers to how models perform across different speakers;
a feat which proves challenging even for the top-performing models investigated here (\cref{subsec:speakers}).
We consider this an important topic which has not been sufficiently investigated for \ac{SER},
though it is long known to impact other speech analysis models~\citep{doddington1998sheep,rajan2021fairness}.

\textbf{Integration of linguistic and paralinguistic streams:}
finally, one of our most intriguing findings is that transformers seem capable of integrating both information streams of the voice signal.
This is evident in how well-performing valence prediction models
retain their effectiveness for synthesised speech lacking emotional intonation (\cref{subsec:tts})
and fail to benefit from fusion with explicit textual information (\cf \cref{subsec:fusion}).
Interestingly, this is only possible when fine-tuning the self-attention layers (\cref{subsec:frozen}),
as keeping them frozen results to complete failure for synthesised speech (\cref{subsec:tts}).
This draws attention to an under investigated aspect of fine-tuning,
namely, how it qualitatively affects the nature of internal representations.
Common understanding sees it as a mechanism through which to obtain better performance,
but our analysis shows that it leads to a fundamental change in how the underlying signal is represented (moving from almost no sensitivity to linguistic content to increased reactivity to it).
This mechanism may be crucial in the pursuit of paralinguistic and linguistic integration
which is key to a holistic understanding of human communication.
However, this integration might prove problematic in cases where the two modalities disagree,
\eg in cases of irony~\citep{burkhardt2017irony}.
Our results also highlight that good valence performance might be language dependent
as models pre-trained on a variety of languages perform worse for valence
compared with comparable models pre-trained only for English (\cref{subsec:valence_gap}).


\section{Conclusion}
\label{sec:conclusion}

Transformers have already revolutionised a very diverse set of artificial intelligence tasks,
including speech emotion recognition.
The present contribution goes beyond previous works
that already established their effectiveness for SER
by conducting a thorough evaluation and analysis of prominent transformer-based speech models
for dimensional emotion recognition.
We obtain state-of-the-art valence recognition performance on {\msppodcast} of $.638$
without using explicit linguistic information,
and manage to attribute this exceptional result
to implicit linguistic information
learnt through a fine-tuning of the self-attention layers.
We release our best performing model ({\wrobustpruned}) to the community~\citep{wagner2022model}.\footnote{{\release}}
Transformer architectures are more robust to small perturbations,
fair on the (gender) group- if not on the individual-level,
and generalise across different domains.
Our findings demonstrate that a new era is dawning in speech emotion recognition:
that of pre-trained, transformer-based foundation models,
which can finally lead to the coveted integration of the two dominant information streams of spoken language, linguistics, and paralinguistics.

\section{\refname}
\printbibliography[heading=none]

\end{document}